# A Set-Theoretic Metaphysics for Wavefunction

Paul Tappenden  paulpagetappenden@gmail.com

1ˢᵗ November 2023

> The lesson to be learned from what I have told of the origin of quantum mechanics is that probable refinements of mathematical methods will not suffice to produce a satisfactory theory, but that somewhere in our doctrine is hidden a concept, unjustified by experience, which we must eliminate to open up the road.
> <div style="text-align:right">(Born 1954: 266)</div>

> May the spirit of Newton's method give us the power to restore unison between physical reality and the profoundest characteristic of Newton's teaching – strict causality.
> <div style="text-align:right">(Einstein 1927: 467)</div>

## ABSTRACT

Set theory brought revolution to philosophy of mathematics and it can bring revolution to philosophy of physics too. All that stands in the way is the intuition that sets of physical objects cannot themselves be physical objects, which appears to derive from the ubiquitous assumption that it's possible for there to exist numerically distinct observers in qualitatively identical cognitive states. Overturning that assumption allows construing the wavefunction of an elementary particle as being a set of *elemental* particles with definite properties. A free electron in an observer's universe is a set of *elemental* electrons on different trajectories, each in an *elemental* "parallel" universe. For any region in an observer's environment which includes part of the *environmental* electron's wavefunction there's a subset of *elemental* electrons located in parallel *elemental* regions. The measure of that subset on the set which is the *environmental* electron is the absolute square of quantum amplitude for the *environmental* region. Decoherence induces Everettian branching as the partitioning of wavefunction into subsets whose measures are the objective probabilities of quasi-classical events within branches. Phase arises through interactions between *elemental* universes, as with Many Interacting Worlds theory, the difference being that an observer's environment is constituted by a *set* of "worlds". That environment contains superpositions as sets of *elemental* particle configurations.

1 Hidden Mechanics

On the 24[th] September 1923 Louis de Broglie made this momentous prediction:

> Le nouveau principe mis à base de la dynamique expliquerait la diffraction des atomes de lumière, *si petit que soit leur nombre*. De plus un mobile quelconque pourrait dans certains cas se diffracter. Un flot d'électrons traversant une ouverture assez petite présenterait des phénomènes de diffraction. C'est de ce côté qu'il faudra peut-être chercher des confirmations expérimentales de nos idées.[1]
>
> (de Broglie 1923: 549, original emphasis)

Notice that he emphasised that he was predicting that *single* particles would diffract (see footnote for a translation). How can the world *be* like that? A century later, there's no consensus on how, and much effort is being made to establish that physics can do without the concept of a mind-independent external world altogether (Müller 2020). Contrariwise, effort continues to be put into preserving realism. Two major themes in that effort are Pilot Wave and Many Worlds theories but there are difficulties in both camps. Pilot Wave struggles to accommodate Quantum Field Theory and Many Worlds struggles with the ontic status of wavefunction, as clearly illustrated in Sean Carroll's recent book when he writes:

> What the World Is Made Of… A quantum wave function
> (Carroll 2019: 49)

> wave functions are superpositions of different possibilities
> (*ibid.*: 64)

> Wave functions may be real, but they're undeniably abstract
> (*ibid.*: 79)

---

[1] This new basic principle of dynamics would explain the diffraction of particles of light, *however small their number*. Further, any moving object could diffract in the right circumstances. A stream of electrons passing through a sufficiently small opening would display diffraction phenomena. It's here that we should perhaps look for experimental confirmation of our ideas. (author's translation)

Concrete objects are undeniably abstract, and made of possibilities? Surely some clarification is necessary! David Wallace is tackling the problem with what he calls a "mathematics-first" approach to the metaphysics of physics, appealing to Ontic Structural Realism, which involves an object-less ontology (Wallace 2022). The project is ongoing and I wish it well; understanding the relation between mathematics and physics is of fundamental importance. However, here I shall be arguing that when it comes specifically to the metaphysics of wavefunction, an ontology of objects bearing definite properties is adequate, so long as objects bearing *indefinite* properties are construed as sets.

Charles Sebens has considered a parallel-universes theory which consists of a set of independently-evolving Pilot Wave universes (Sebens 2015: §3). Everettian *branching* is re-interpreted as the partitioning of the set of Pilot Wave universes into subsets (Everett 1957: 459). What Sebens demonstrates in the subsequent section is that the set of independently evolving Pilot Wave universes could be replaced by a set of interacting universes with the interactions mediated by "Newtonian" forces. So, a mixed particle-and-wave ontology is replaced by a particle-only ontology; the wave aspect of unitary evolution being generated by particle interactions, like waves generated in a vibrating mass of sand. Similar ideas are to be found in (Hall *et al*. 2014; Boström 2015). For Many Interacting Worlds theory, like Pilot Wave, microscopic objects always have definite physical properties. Those objects are generally taken to be particles but similar ideas might apply to field values, though I shall speak in terms of particles for the sake of convenience.

Another current variant on parallel-universes theories is the Quantum Modal Realism proposed by Alastair Wilson and derived from the version of the concept of *divergence* introduced by Simon Saunders (Wilson 2020; Saunders 2010: 196-7). That's in the tradition of Everetttian "pure wave" theories; the individual universes don't contain particle trajectories and interactions between them are not invoked in order to explain the partitioning of sets of individual universes into subsets constituting Everettian branches. As with other versions of Many Worlds which lack hidden-variables, the ontology of wavefunction remains problematic.

In what follows I shall attempt to succinctly present key ideas which have been developed in a series of papers; clarifying, correcting and adding further thoughts (Tappenden 2017, 2021, 2022, 2023a, 2023b). I shall argue that what we should eliminate "to open up the road" is the assumption that sets of physical objects are not themselves physical objects. An *environmental* two-slit apparatus could be a set of *elemental* apparatuses which are macroscopically isomorphic but microscopically anisomorphic, in each of which a single *elemental* particle passes through a single *elemental* slit.

The term *elemental* is to be understood in a strictly set-theoretic sense. I must emphasise that the term *elemental particle* as I shall be using it has a very different sense from the common term *elementary particle*. An *environmental* electron is a single elementary particle which is to be thought of as a set of *elemental* particles. The idea is that any object in an observer's environment is a set, whether it's a particle, an atom, the observer's body or a galaxy. So, an *environmental* galaxy is a set of *elemental* galaxies and an *environmental* atom is a set of *elemental* atoms.

An observer exists in an *environmental* universe which is a set of *elemental* universes but an *environmental* universe can exist without observers existing within it. That was the case for our universe before life began. An object is *environmental* if it's the sort of object which could exist in an observer's *environmental* universe and an *environmental* object is always a set of *elemental* objects. Bearing this in mind, the concepts of *environmental* and *elemental* objects could be summarised as:

> *Environmental* Object
>
> The sort of object which could exist in an observer's environment. Always a set of *elemental* objects.
>
> *Elemental* Object
>
> The sort of object which could be a set-theoretic element of an *environmental* object. Always having definite values for physical properties. Not all sets of *elemental* objects are *environmental* objects.[2]

The dichotomy between *environmental* objects and *elemental* objects will be essential, so I'll italicise those terms throughout. A single electron in an observer's *environmental* universe can be a set of *elemental* electrons, each following a different trajectory in an *elemental* universe.

---

[2] To jump ahead, the overall idea is that the *environmental* universal wavefunction is a set of *elemental* universes and its partitioning is Everettian branching. So it's the decoherence involved in unitary evolution which selects the subsets of *elemental* objects which are *environmental* objects.

If the two-slit apparatus is a set of apparatuses it follows that the observer's body is a set of *doppelgängers*. The observer *splits* Everett-style into observers making different electron-detection observations because their body partitions into cerebrally anisomorphic bodies which are subsets instancing those different observations. Eliminating the idea that sets of physical objects are not themselves physical objects requires eliminating the idea that there's a one-to-one relation between observers and doppelgängers. The core conceptual change required is to drop the ubiquitous assumption enshrined in Hilary Putnam's seminal *Twin Earth* thought experiment, which has had a profound influence on contemporary analytic philosophy of mind for nearly half a century (Putnam 1975). So, the next section is on metaphysics of mind, to prepare the way for metaphysics of physics, followed by a conclusion and its consequences.

2 Mind in a Multiverse

The set-theoretic metaphysics for wavefunction involves a set of universes, a multiverse. They are to be thought of as distributed in configuration space. A difference from classical statistical mechanics is that configurations constitute a space of actualities, rather than a fictional space of possibilities. *Environmental* objects are set-theoretically extended in configuration space, so to speak.

Fundamental to making sense of this is our conception of how observers are *situated* in a multiverse, to use Jeffrey Barrett's term (Barrett 2021). The metaphysics has to be consistent with observed phenomena and phenomena depend on how observers are situated. A useful step in thinking about how an observer is situated in a multiverse is to temporarily set aside quantum theory and consider what Max Tegmark has called the Level I multiverse (Tegmark 2007).

According to current cosmology there's no evidence that space is finite. Our so-called *observable universe* surrounds us out to a distance of about 46 billion lightyears. That's the maximum distance in our absolute elsewhere of objects which we observe now as they were just after the Big Bang. Beyond that, we have no reason to believe that the overall pan-galactic structure changes from place to place, so there can be duplicate observable universes. Tegmark has estimated that the average separation of universes observationally isomorphic to ours is $10^{10^{117}}$ metres. So far as we know, our observable universe may be one of a denumerably infinite number of observationally identical copies.

There's a metaphysical bias in that last paragraph. Is *our* observable universe one among many, or is our universe somehow the very plurality itself? If the plurality, or the mereological sum, it would have infinite mass, which isn't the case. However, our observable universe could be the *set* of individual universes so long as we relinquish the idea that sets of physical objects are necessarily abstract.

Willard Van Orman Quine took a first step in that direction by suggesting that *individual* physical objects could be interpreted as sets that are their own sole elements, which have come to be known to logicians as *Quine atoms* (Quine 1969: 31). For more on Quine's assimilation of individuals to set theory see (Tappenden 2022: §4). The further step required is to adopt this hypothesis:

Concrete Sets

A set of physical objects which are Quine atoms has all and only the properties which its elements share, other than those that are logically excluded.

Excluded properties include number of elements and value-definiteness. Note that this doesn't entail that sets of sets of concrete Quine atoms are themselves concrete; we can continue to think of those as being abstract, except in the case of Quine atoms themselves, which are identical to their unit sets.

Adopting that hypothesis allows us to construe our *environmental* universe as a set of *elemental* universes and any object in an observer's environment to be set of *elemental* objects which are Quine atoms. All *elemental* objects are Quine atoms. I should stress that the term *Quine atom* has nothing to do with size. *Elemental* electrons are Quine atoms and *elemental* galaxies are Quine atoms.

This has surreal consequences. For instance, the set of the elements of an *environmental* sapphire and of an *environmental* emerald of equal mass is a set of Quine atoms which is a third gem with half the mass of the pair of *environmental* gems and *indefinite* colour, since its elements don't have the same colour. How could that possibly be the case? Where would the extra mass *be*? A thought experiment helps.

Consider two isomorphic *environmental* cubicles containing doppelgängers beside tables on which are matched boxes. One contains the sapphire, the other the emerald. The doppelgängers both emit noises which sound like the utterance "I see one box". Conventionally, the noises would be interpreted as tokens of two utterances by two distinct observers, each referring to one of the boxes, but an alternative interpretation is possible, *given* Concrete Sets.

There's a single observer who makes a single utterance tokened by the set of *elemental* vocal noises. That observer refers to the box which is the set of the *elemental* boxes containing the elements of the sapphire and the emerald. That observer's box contains a gem with *indefinite* colour and position because it has subsets with different colours and positions. The extra gem is nonetheless a massive object in that observer's environment. The missing mass exists in the environment of the missing observer, so to speak, the *single* observer who's missing when the observer-doppelgänger relation is assumed to be one-to-one, as in Putnam's Twin Earth thought experiment, which I shall outline in a moment. The alternative is one-to-many:

> Unitary Mind
>
> The mentality of a single observer may be instanced by many, relevantly isomorphic, brain-like objects.

Concrete Sets and Unitary Mind are interdependent. Each is required to render the other intelligible.

2.1 The External World

Putnam profoundly challenged the idea that mentality is wholly determined by cerebral processes, which had become known as mind-brain identity theory in the 1960s. He put it in doubt that there's a dichotomy between *internal* mental states determined by localised processes going on in brain-like objects and an *external* world where all physical processes take place.
    The thought experiment is simple and elegant. All sorts of quibbles are possible, but here's the *eureka* idea:

> Twin Earth
>
> Oscar and Toscar have isomorphic bodies. Oscar's on Earth where all the liquid in the lakes and lagoons is water and nobody has yet discovered that water is $H_2O$. Toscar is on Twin Earth, which is isomorphic to Earth except that all the liquid in the lakes and lagoons is twater and it's yet to be discovered that its molecular constitution is XYZ. When Oscar thinks about water he thinks about stuff which is $H_2O$ and when Toscar thinks about twater he thinks about stuff which is XYZ, so they're entertaining different thoughts whilst having bodies which are isomorphic.

Quibbles aside, grant Putnam his insight. The argument is clearly undermined by Unitary Mind. Replace Oscar and Toscar by Scar and he thinks about stuff which has indefinite molecular constitution. That removes Putnam's objection. We are free to return to the hypothesis that the contents of mental states are indeed wholly determined by processes going on in brain-like objects, known as *mental content internalism*. All that's required is to allow that the mind-brain relation should no longer be thought of as one-to-one and physical objects in an observer's environment may be sets of physical objects.

2.2 The View from Somewhere

When Scar pioneers the chemical analysis of a vial of the liquid in lakes and lagoons, the pair of doppelgängers make parallel movements in Putnam's multiverse. As the analysis proceeds the doppelgängers are exposed to different sensual stimuli and so emit different noises, one sounding like "Ah, $H_2O$!" and the other sounding like "Ah, XYZ!" We can interpret what's happened as Scar having split, Everett-style, into two observers making different observations. This suggests a conceptual connection between set theory, Everettian fission and the existence of objects with indefinite properties in quantum theory.

That's an alternative third-person view of how observers are situated in a multiverse inhabited by doppelgängers, but how does it look from the first-person point of view, assuming that a denumerably infinite number of copies of our observable universe do indeed exist? Isn't the concept of space itself somehow undermined?

No. All that's changed is our conception of how the universe beyond our observable universe is constituted. From the Unitary Mind point of view, beyond the observable universe is a massive macroscopic "superposition" of every way things have evolved since the Big Bang. That must be so because what we refer to as our observable universe is to be thought of as a set of universes which are isomorphic but surrounded by different distributions of matter and energy. Even without quantum theory, the concept of a spatially infinite universe coupled with Unitary Mind brings a fundamental dichotomy to material existence *relative to observers*. *Environmental* objects are sets of *elemental* objects.

From the god's eye "view from nowhere" in the Level 1 multiverse there's no such dichotomy, all is *elemental*. But for observers who inhabit sets of *elemental* universes the dichotomy is real: *environmental* and *elemental* physical objects are fundamentally different in that *environmental* objects can have indefinite values for physical properties such as position and momentum whereas *elemental* objects cannot.

I should mention that Nick Bostrom has briefly considered and rejected Unitary Mind, which he calls *Unification* (Bostrom 2006: 186). He endorses the usual intuition about the mind-body relation when he writes:

> It would, to say the least, be odd to suppose that whether one's own brain produces phenomenal experience strongly depends on the happenings in other brains that may exist in faraway galaxies that are causally disconnected from our solar system
>
> (*ibid*.: 188)

That doesn't take into account that the alternative interpretation of the mind-body relation entails an alternative interpretation of the constitution of *environmental* objects. According to the set-theoretic metaphysics the *elements* of an observer's environment don't *need* to be causally connected, they could, from a god's eye view, be multiple copies of island universes scattered through space.

When it comes to adding quantum theory to a Level I multiverse, think about the decay of an unstable nucleus and suppose, for the sake of argument, that the process is *stochastic*. An *environmental* particle is an infinite set of *elemental* particles, each of which has a *propensity* to decay given by its half-life. Necessarily, since the set is infinite, after the half-life has elapsed the original set of particles has partitioned into decayed and undecayed subsets of equal measure. An observer observing the *environmental* nucleus after one half-life would fission into observers seeing and not seeing decay, and the subset measures of the downstream observers' bodies would necessarily be equal because that's the *probability measure*. The observer fissions because their body partitions. The *environmental* nucleus is continually partitioning into undecayed and decayed subsets, on a timescale of $10^{-23}$s., with the measure of the former decreasing and that of the later increasing until the measures are equal after one half-life.

Note that the process of fission, both of observers and *environmental* objects is necessarily stepwise. Fission cannot be a continuous process since that would imply an infinite number of fissions between any two, however close. An unstable nucleus cannot fission more rapidly than the light-time of its radius, an observer cannot fission more rapidly than the duration of an observation, which corresponds to the characteristic timescale of the cerebral processes which are instances of observations, a substantial fraction of a second in the case of humans.

According to Unitary Mind, an observer who is well-informed about their situation in a stochastic Level I multiverse, and who is about to observe an unstable particle one half-life after its creation, should expect to fission into two observers making different observations in two distinct environments whose probability measures relative to the original environment are 0.5. What should the original observer *expect to observe*? This raises questions about persistence and probabilistic expectation which need to be resolved. Once that's done, we can go on to replace the concept of stochastic quantum processes with that of *dendritic* processes, as suggested by Everett (*op.cit.*: 460). A dendritic process being like a stochastic process except that so-called possible outcomes with fractional probabilities are replaced by actual outcomes with fractional probabilities.

2.3 Fission and Persistence

Independently of quantum theory, the problem posed by the fission of observers was extensively discussed by Derek Parfit, who concluded that the concept of continuing identity must be abandoned in such situations (Parfit 1984: 199ff.). In that case, our observer who expects to fission has no warrant for expecting that *they* will observe anything at all! If Scar fissions into Oscar and Toscar, and Oscar is not Toscar, then Oscar *and* Toscar can't *both* be Scar. So, if there's no reason to believe that *either* Oscar *or* Toscar is Scar, there's no reason to believe that Scar will persist beyond fission.

A way out of this impasse was introduced by Ted Sider as *Stage Theory* (Sider 1996; 2001: 201; Hawley 2001: Ch.2). It was first explicitly applied to Everettian theory in (Tappenden 2008: 313) and implicitly by Hilary Greaves (Greaves 2004: §4.1.1). The idea is that persisting observers, at any given moment, are temporal parts of their histories and bear *temporal counterpart* relations with the other temporal parts. Scar *will be* Oscar and *will be* Toscar. Oscar *was* Scar and Toscar *was* Scar. Three distinct individuals bear the relations *was* and *will be* relative to each other.

Stage theory applies to objects as well as observers. The temporal counterpart relation is mediated by spatiotemporal continuity in the case of objects. Since observers are related to objects in the sense that their cognitive states are instanced by brains, observers' persistence involves both spatiotemporal and cognitive continuity.

The liquid in Scar's vial has indefinite constitution because it's a set of two vials which contain liquids with different molecular constitutions. A little later, Oscar's vial contains $H_2O$ and Toscar's vial contains XYZ. Oscar's and Toscar's observations occur in two distinct *environmental* regions which are spatiotemporally continuous with the *elements* of the *environmental* region where Scar makes his pre-measurement observation.

Scar's *environmental* brain is a set of two isomorphic *elemental* brains; its post-measurement future temporal counterpart is a set of two *anisomorphic* brains, one registering $H_2O$ and the other XYZ. On the assumption that contradictory observational states can't exist, that entails that Scar fissions into an Oscar observing $H_2O$ and an Toscar observing XYZ. Concrete Sets together with Unitary Mind in Putnam's multiverse entail that Oscar's and Toscar's *environmental* vials are Quine atoms; they have no elements other than themselves. So the dichotomy between *environmental* and *elemental* objects breaks down in Putnam's idealised multiverse, where vials are presumed to be individuals.

Stage theory has remained marginal amongst theories of persistence, as discussed in (Tappenden 2022: §2.1). But there's no known reason to reject it, and it's tailor-made for dealing with persistence through fission:

> Stage Theory
>
> A persisting observer or object at time *t* is a temporal part of its history and the other temporal parts are its *temporal counterparts*. It bears the relation *was* to temporal counterparts at times earlier than *t* and *will be* to temporal counterparts at times later than *t*.

Returning to a Level 1 stochastic multiverse and the observation of an unstable particle after one half-life, a single observer, knowingly about to fission, can expect to be two different observers making different observations in different environments, each with probability measure 0.5. How is the observer to make sense of that? Only via the concept of probabilistic expectation, which requires that the pre-fission observer should be uncertain about their future. But how can they be uncertain about their future if certain that they will fission?

2.4 Fission and Uncertainty

The concept of stochasticity involves *objective* probability in the sense that if the decay of an *elemental* unstable particle is stochastic then it's a mind-independent property of that particle that it's probability of decay after one half-life is 0.5. So, if an observer's *environmental* unstable particle is an infinite set of *elemental* stochastic particles, the objective probability of decay of an *environmental* unstable particle for one half-life *is* the measure of the decayed subset after one half-life: 0.5, the probability measure. If that observer observes the state of an *environmental* unstable particle after one half-life life they fission into an observer observing decay and an observer not observing decay and the subset measures of the two downstream observers' environments are 0.5 relative to the environment of the upstream observer.

Uncertainty about the future is a cognitive state which involves the assigning of *subjective* probabilities to future observations, otherwise known as degrees of belief or *credences*. It seems rational that an observer who believes that the objective probability of a future event is *p* should assign a *credence* equal to *p* for the future observation of that event. David Lewis dubbed that thought the *Principal Principle* (Lewis 1980: 266).

Lewis introduced the term in the context of considering the radioactive decay of tritium, which he interpreted as a stochastic process. In that context, an observer assigns credences to *possible* future observations only one of which will actually occur. So Lewis's *stochastic* Principal Principle requires a caveat; it only applies if the observer doesn't have access to *inadmissible* evidence. If the observer knows in advance what the outcome of a stochastic event will be then it's rational to assign a credence of 1.0 to the future observation of that outcome. For more on Lewis and *admissibility* see (Tappenden 2021: §2.2) and references therein.

However, from the point of view of an observer whose *environmental* unstable particle is an infinite set of *elemental* stochastic particles no such caveat is required. The objective probabilities of future observations are the subset measures of the post-measurement environments where the downstream observers will make those observations. The well-informed upstream observer knows what the objective probabilities are and there's nothing more to know about what will occur; all the measurement outcomes will occur and the observer will fission. So Lewis's original term needs a qualification:

Dendritic Principal Principle

For an observer who believes that the objective probability
for the occurrence of future event *E* is *p*, it's rational to
assign credence *p* to the future observation of *E*.[3]

Our observer about to observe an *environmental* unstable particle after one half-life should thus assign a credence of 0.5 to observing decay and a credence of 0.5 to not observing decay. Our observer is uncertain about what they will *observe* just like an observer in a single stochastic

---

[3] The Deutsch-Wallace decision-theoretic argument involves a detailed analysis of the rational principles involved here (Deutsch 1999; Wallace 2012: Ch.5)

universe. An observer in a Level 1 stochastic multiverse who believes that the observer-doppelgänger relation is one-to-one and who believes that they're observing a stochastic process is deceived, according to Unitary Mind; what the observer is *in fact* observing is a dendritic process, one where all outcomes occur, whatever their objective probabilities.

That can seem odd because our single observer must be, at the same time, *certain* that the *environmental* particle will be a particle which has decayed and will be a particle which has not decayed since the objective probability for that outcome is 1.0, the sum of the measures of the undecayed and decayed subsets. But there's no contradiction here; the fact that both outcomes will occur entails that each will occur. Our observer is uncertain as to what they will *observe* whilst certain as to what will *occur*. Our observer is certain that the *environmental* particle will be a decayed *environmental* particle even thought the objective probability of decay is 0.5. That's just a logical consequence of the fact that the environmental particle will be an undecayed *environmental* particle *and* will be a decayed *environmental* particle.

We've been considering Unitary Mind in a Level I stochastic multiverse, but stochasticity is not for us. In the spirit of Albert Einstein's sentiment in the opening quote, Everett sought to replace the concept of stochastic process with that of a *deterministic* dendritic process. He wrote:

> The theory based on pure wave mechanics is a conceptually simple, *causal* theory
>
> (*op.cit.*: 462, my emphasis)

Everett had a truly revolutionary idea, but there's still no agreement amongst Many Worlds theorists about what a fully satisfactory version of the theory is. Everett put the cat amongst the pigeons of everyday intuition. What's being argued here is that the theory should be *both* deterministic *and* objectively probabilistic. The absolute square of amplitude of the Everettian branches which issue from dendritic quantum processes just *is* objective probability. That's because the Born rule, as applied to stochastic theory, assigns objective probabilities to possible outcomes and Everett's *eureka* idea was to replace the concept of multiple possible outcomes with that of multiple actual outcomes. We need to abandon Everett's assumption that dendritic processes are not objectively probabilistic; furthermore, we need to abandon his assumption that the theory should involve a pure wave mechanics, lacking hidden-variables.

2.5 Recap

Adopting the Concrete Sets hypothesis allows observers to be situated in sets of parallel universes rather than individual universes. Unitary Mind entails the fission of observers and *environmental* objects in a Level 1 stochastic multiverse. Stage Theory enables observers and objects to persist through fission. The Dendritic Principal Principle induces a rational observer to be uncertain about their future observations whilst certain as to what will occur. We are thereby conceptually equipped to think about observers in a different sort of multiverse. The Level 1 *cosmological* multiverse may well exist, if so, it adds a further layer of structure to what follows, but best to set it aside for now and think about a multiverse where *environmental* quantum processes are dendritic and *elemental* quantum processes are deterministic, not stochastic.

3 The Quantum Multiverse

In the set-theoretic quantum multiverse the observer's *environmental* universe becomes a set of *elemental* universes which includes all possible particle configurations consistent with observations; in particular, observations of Born rule probabilities. What a wavefunction *is* is an interacting set of *elemental* particle configurations, each in an *elemental* universe. Like the Cheshire cat, the *environmental* free electron teases the observer with its grin; what exists in the observer's environment is a set whose elements are elsewhere, in configuration space.

The configurations are on the position basis, in the sense that an *environmental* particle only has a definite *environmental* position when its elements are at corresponding *elemental* positions. Correspondence of positions between *elemental* universes is determined by the distribution of macroscopic objects in an observer's past lightcone at time *t*, since they all have definite *environmental* positions. An observer at time *t* inhabits a set of "parallel" universes in the sense that the macroscopic contents of the surface of the observer's past lightcone is a set of congruent macroscopic configurations. Note that this doesn't constrain macroscopic objects to have definite *environmental* positions in an observer's absolute elsewhere.

Consider the wavefunction of an *environmental* electron arriving at an *environmental* detector array after having passed through a two-slit apparatus. The absolute square of amplitude of the wavefunction yields the probability that a particular *environmental* detector will be observed to fire. The reason that's so, according to the set-theoretic metaphysics, is that the *environmental* detector array is a set of *elemental* detector arrays for each of which an *elemental* electron following a trajectory is about to impact an *elemental* detector. And the measure of the subset of *elemental* arrays where a particular *elemental* detector is about to be impacted is the probability that the corresponding *environmental* detector, which is the set of those *elemental* detectors, will be observed to fire.

Talk of probabilities here presumes measures on *infinite* sets. Are *environmental* objects infinite sets of *elemental* objects? That would seem to depend on whether spacetime is continuous. However, even if it's not, and the relevant configurations are finite in number, the possibility of the cosmological multiverse can be invoked until such time as we have evidence that space is finite. That provides a denumerably infinite number of *elemental* objects for every configuration.

On observing the impact of the *environmental* electron on the *environmental* array the observer's *environmental* body partitions into subsets of *elemental* bodies, each subset being the body of an observer observing the firing of a different post-impact *environmental* detector. The environment of the pre-impact observer partitions into the environments of the many post-impact observers and the subset measures of those environments relative to the pre-impact environment are the probabilities of the observations of particular post-impact *environmental* detectors firing. The interference pattern for a single *environmental* electron arriving at an *environmental* array is the distribution of measures of the subsets of *elemental* arrays corresponding to each detection site. When the *environmental* electron passes through the *environmental* slits there are two subsets of *elemental* electrons each of which passes through one *elemental* slit. Following Many Interacting Worlds theory, it's the interactions between the *elemental* electrons in those two subsets which gives rise to the interference pattern. We observe phenomena at the *environmental* level; the mechanism giving rise to those phenomena operates at the *elemental* level.

That mechanism is manifest as the partitioning of wavefunction into subsets. The wavefunction of the two-slit apparatus partitions into as many Everettian branches as there are detectors in the array (plus one branch where no detector fires). The microscopic superpositional state of the *environmental* electron arriving at the detector array is amplified so as to create a macroscopic superposition, each component of which is a subset of *elemental* detector arrays. In each branch an observer sees a different detector fire (or none). The process of amplification is mediated by decoherence, as explained in (Wallace 2012: Ch.3).

An *environmental* electron always has indefinite spin relative to some orientations but an *elemental* electron cannot, since all *elemental* objects are in definite states. That being so, any *elemental* electron must simply be either spin-up or spin-down relative to some particular orientation[4]. Given Concrete Sets, an *environmental* z-spin-up electron must have a subset of *elemental* electrons with definite spins relative to the z-axis and all the *elemental* electrons in that subset must be spin-up, otherwise the *environmental* electron wouldn't have definite z-spin. Furthermore, the *environmental* electron must have a subset of *elemental* electrons for every other orientation and each of those subsets must have subsets of spin-up and spin-down *elemental* electrons since the *environmental* electron has indefinite spin relative to those orientations. Necessarily, the subset measures of spin-up and spin-down *elemental* electrons for a given orientation must be the probabilities for the *environmental* z-spin-up electron to be measured spin-up or spin-down relative to that orientation.

David Deutsch once wrote:

pilot-wave theories are parallel-universes theories in a state of chronic denial.
(Deutsch 1996: 226)

Those words close a paragraph where he describes Pilot Wave theory as only filling one of the "grooves" in an "immensely complicated multi-dimensional wavefunction" and claims that "the 'unoccupied grooves' must be physically real". A retort could be that parallel-universes theories have been hidden-variable theories in denial. For Deutsch's unoccupied grooves to be physically real they need to be occupied. The set-theoretic metaphysics fills *all* the grooves by making the *environmental* wavefunction a set of interacting *elemental* objects.

Deutsch's vision of the quantum multiverse as a partitioning set of parallel universes is similar to the quantum multiverse which I've described (Deutsch 1985: 20). But there were problems for Deutsch's 1985 view. First of all, the individual parallel universes are not supposed to be stochastic, so why does the set partition?[5] In his popular book *The Fabric of Reality*, he hints at interaction between individual universes (Deutsch 1997: 43-4). The other problem was the provision of a preferred basis to select the set of universes. Sara Foster and Harvey

---

[4] Sebens introduces this idea in the context of his version of Many Interacting Worlds theory (*op.cit.*: §13).

[5] The same question arises for Wilson's diverging set of parallel universes (*op.cit.*).

Brown showed that there was a fatal flaw (Foster & Brown 1988). The set-theoretic metaphysics provides the position basis via hidden-variables.

4  Conclusion and Consequences

Set theory brings a new "dimension" to physics, which Michael Lockwood gestured towards as a *Pickwickian* dimension, attributing the idea to Deutsch (Lockwood 1989: 232). I've argued that *environmental* objects are set-theoretically "extended" in configuration space, a space of actual configurations. An *environmental* free electron is a set of *elemental* electrons on different trajectories. A macroscopic *environmental* object is a set of macroscopic *elemental* objects constituted by different configurations of particles (and/or fields). The body of an observer is a set of doppelgängers which partitions into cognitively distinct subsets in measurement contexts, so the pre-measurement observer bears the relation *will be* to each of the post-measurement observers making different observations.

Odd as it may seem, there's no reason for this to make much difference to how we treat probabilities in everyday life, though there might be exceptional circumstances which do make a difference (Papineau & Rowe 2023). Different futures have probabilities, as before, they're just concrete actualities rather than abstract possibilities. Actors will strive to make the futures they desire more probable. The notion of free will remains as challenging as it was in classical, deterministic-but-not-probabilistic physics.

However, there's an elephant in the room of Many Worlds theory which shouldn't be ignored, going by the names of *quantum suicide* or *quantum Russian roulette*. Theorists have blown hot and cold on the subject over the last few decades (Tegmark 1997, Carroll *op.cit.*: 209). For a quantum Russian roulette setup, the set-theoretic metaphysics has it that there's a survival branch with an objective probability of 5/6 and the person pulling the trigger can be sure that they will be a person who has survived, however many times they repeat the operation.

David Lewis concluded:

You who bid good riddance to collapse laws, you quantum cosmologists, you enthusiasts of quantum computing, should shake in your shoes. Everett's idea is elegant, but heaven forefend that it should be true!

(Lewis 2004: 21)

For quantum statistical mechanics with actualised configuration space every erstwhile "possible" physical future has some probability, be it ever so small. There are always survival branches, even when you hit the deck after falling from an aircraft without a parachute, or when you're lingering in a hospice with "terminal" cancer. "Aye, there's the rub". Does some sort of eternal *terminal limbo* await us all, rather than oblivion? Coping with that prospect is a challenge we may have to face.

## Acknowledgments

Heart and soul felt thanks to Devin Bayer for inestimable help with preparing this draft, albeit that I know he's still not convinced that it works. Thanks also to Jørn Klovfjell Mjelva and Simon Saunders for feedback which brought significant changes, even to the title! Further issues, better left out of what's meant to be an effective introduction, are discussed in (Tappenden 2023b §§ 2.3, 3.1, 3.2)